\documentclass[useAMS,usenatbib]{mn2e}
\usepackage{amsmath} 
\usepackage{subfigure}
\usepackage{graphicx}
\usepackage{times}%
\usepackage{natbib}
\usepackage{lscape}
\usepackage{psfig,natbib}
\usepackage{deluxetable}

\def\aap{A\&A}
\def\apj{ApJ}

\def\apjl{ApJ}
\def\mnras{MNRAS}
\def\araa{ARA\&A}
\def\aj{AJ}

\def\nat{Nat}

\def\pasp{PASP}

\title[Chance coincidence of a GRB with  galaxy on sky]
{Probability for chance coincidence of a gamma-ray burst with a galaxy on 
the sky}
\author[M. A. Campisi and L.-X. Li]
{Maria Angela Campisi\thanks{E-mail: campisi@mpa-garching.mpg.de} and
Li-Xin Li\\
Max-Planck-Institut f\"ur Astrophysik, Karl-Schwarzschild-Strasse 1, 
D-85748 Garching bei M\"unchen, Germany
}

\begin{document}
\date{Accepted 2008 September 11. Received 2008 September 11; in original form 2008 June 18}
\pagerange{\pageref{firstpage}--\pageref{lastpage}} \pubyear{2007}
\maketitle
\label{firstpage}

\begin{abstract}
The nearby long GRB 060614 was not accompanied by a supernova, challenging 
the collapsar model for long-duration GRBs and the traditional classification
scheme for GRBs. However, Cobb et al. have argued that the association of 
GRB 060614 and its host galaxy could be chance coincidence. In this work we
calculate the probability for a GRB to be randomly coincident with a galaxy 
on the sky, using a galaxy luminosity function obtained from current galaxy
surveys. We find that, with a magnitude limit that current telescopes can 
reach and an evolving galaxy luminosity function obtained from VVDS, the 
probability for chance coincidence of a GRB with a galaxy of redshift $<1.5$
is about several percent. These results agree with previous estimates based on
observed galaxies. For the case of GRB 060614, the probability for it to
be coincident with a $z<0.125$ galaxy by angular separation $<0.5^{\prime
\prime}$ is $\approx 0.02\%$, indicating that the association of GRB 060614
and its host galaxy is secure. If the telescope magnitude limit is 
significantly improved in future, the probability for GRB-galaxy association
will be considerably large, making it very problematic to identify a GRB 
host based only on the superposition of a GRB and a galaxy on the sky.

\end{abstract}
\begin{keywords}

gamma-rays: bursts -- galaxies: luminosity function, mass function.

\end{keywords}


\section{Introduction} \label{sect:intro}

Observation of host galaxies of gamma-ray bursts (GRBs) is very important
for understanding the nature of GRBs. Current observations reveal that
long duration GRBs occur in star-forming galaxies, consistent with the 
general belief that long GRBs are produced by the death of massive stars 
\citep[and references therein]{con05,fru06,tan07,wai07}. The discovery of 
the connection between long GRBs and core-collapse Type Ibc supernovae 
\citep[and references therein]{gal98,li06,woo06b} supports the collapsar 
model of long GRBs (MacFadyen \& Woosley 1999; MacFadyen, Woosley \& 
Heger 2001).

In contrast, short-duration GRBs are found in both early- and late-type 
galaxies, similar to the situation of Type Ia supernovae. The rate of star 
formation in the host galaxies of short GRBs is often lower than that in the 
hosts of long GRBs \citep[and references therein]{ber06}. So far no supernovae
have been found to be associated with short GRBs.

The difference in the observed host properties for short and long GRBs 
supports the idea that short and long GRBs have different progenitors. Long 
GRBs are believed to arise from the death of massive stars (the collapsar
model)---most likely the Wolf-Rayet stars since all observed supernovae 
associated with GRBs are Type Ic, while short GRBs are more likely produced 
by the merger of compact stars---neutron star-neutron star merger and black 
hole-neutron star merger (Li \& Paczy\'nski 1998; O'Shaughnessy, Kalogera 
\& Belczynski 2007).

However, the above scenario is challenged by the observation of GRB 060614.
This is a long burst with a duration $\sim 100$~s, with a host galaxy at 
redshift $z=0.125$ \citep{del06,fyn06,geh06}. For a long GRB that has such 
a low redshift it is expected that a supernova associated with it should be 
observed. However, despite extensive observation on its host, no supernova 
has been found down to limits fainter than any known Type Ic SN and hundreds 
of times fainter than the archetypal SN 1998bw that accompanied GRB 980425.
This challenges the ordinary GRB classification scheme based on GRB
durations and the general belief that long GRBs are produced by the 
core-collapse of massive stars \citep{zha06,wat07}.

In fact, except its duration, GRB 060614 is much like a short GRB in many
aspects. Besides the fact that it has no associated supernova, GRB 060614
has a vanishing spectral lag that is typical of short GRBs \citep{man07}. 
Its lightcurve has a very hard and short-duration initial peak, followed by 
an extended soft emission. \citet{zha07} have shown that if this burst had
an eight time smaller total energy, it would have been detected by BATSE as 
a marginal short-duration GRB, and would have properties in the {\em Swift} 
BAT and XRT bands similar to GRB 050724.

GRB 060505 has a duration $\sim 4$ s and a host galaxy at $z=0.089$. No
supernova has been detected at the location of this burst also 
\citep{fyn06,ofe06}, so GRB 060505 has also been considered as a long
GRB without a supernova. It has been argued that GRB 060505 is indeed a short
burst 
\citep[see, however, Th\"one et al. 2008; McBreen et al. 2008]{ofe06,lev07}.
Models for ``long GRBs without supernovae'' have been proposed (King, Olsson
\& Davies 2007).

On the other hand, it has been suggested that GRB 060614 and its host
galaxy was just a coincidence rather than a physical association 
\citep{cob06b}. By counting the number of galaxies observed by SMARTS in a 
field centered on the burst, \citet{cob06b} showed that the probability for a 
chance superposition of GRB 060614 and a galaxy along the line of sight is 
$\sim 1\%$. This probability is high enough to cause that several cases of 
chance superposition may have happened for {\em Swift} GRBs. This conclusion
is enforced by a more detailed study by \citet{cob08}.

The results of Cobb et al. have raised an important question in identifying
GRB host galaxies based only on the superposition of a GRB and a galaxy on the
sky \citep{leva07}. For a telescope with very high sensitivity, it would 
observe many 
galaxies on the sky, then the probability for a GRB to be aligned with a 
galaxy could be high. Then, unavoidably, some GRB hosts discovered in this
approach might be superficial, i.e. they are not physically related to the
GRBs. Cobb et al. obtained their results by using galaxy survey data. It is
interesting to verify these results with an independent, more theoretical
approach.

In this paper, we calculate the probability for a GRB to be coincident with 
a galaxy on the sky using galaxy luminosity functions and compare the results
with that of Cobb et al. obtained with different ways. Then, we use our
results to assess at what a level we can trust the GRB host galaxies that 
have been found so far. The UVOT on {\em Swift} can resolve a source to
sub-pixels ($\sim 0.2^{\prime\prime}$). Hence, in our calculations we regard
a GRB as a point source.

The approach that we adopt has the benefit of extending beyond the limit
of current surveys and to broader types of problems. For example, with 
slight modification it can be applied to the calculation of the probability
of Ly$\alpha$ forests in the spectra of quasars and GRBs which has important
applications in probing the high-$z$ Universe \citep{loe02}.


\section{The Galaxy Luminosity Function} \label{sect:lum}

The galaxy luminosity function (LF) is a fundamental characteristics of the 
galaxy population and is essential for studying statistical properties of 
galaxies and their evolution. It gives the abundance of galaxies 
as a function of their luminosity, defined by the comoving number 
density of all galaxies with luminosity between $L$ and $L+dL$ at redshift 
$z$. The LF of a population of galaxies is usually described by the Schechter 
function (Schechter 1976)
\begin{eqnarray}
	\Phi(L)dL=\Phi^* \left(\frac{L}{L^*}\right)^\alpha \exp 
		\left(-\frac{L}{L^*}\right) \frac{dL}{L^*} \;,
	\label{lum}
\end{eqnarray}
where $L^*$ is a characteristic luminosity, the constant $\alpha$ is the 
faint-end slope, and $\Phi^*$ is the normalization. These three parameters 
are determined by fitting the LF to the data from a galaxy survey.

The LF is often expressed in terms of magnitudes rather than luminosities, 
which is more convenient to use in UV and optical observations. The absolute 
magnitude $M$ is related to the galaxy luminosity by $M-M^*=-2.5 \log 
(L/L^*)$, where $M^*$ is a characteristic magnitude corresponding to the 
characteristic luminosity $L^*$. Then, the Schechter LF becomes
\begin{eqnarray}
	\Phi(M)dM ~=~ (0.4  \ln 10) \Phi^* \hspace{4cm} \nonumber\\
		\times 10^{0.4 \left(\alpha+1\right)\left(M^*-M\right)} 
		\exp\left[-10^{0.4(M^*-M)}\right] dM \;. 
	\label{lum2}
\end{eqnarray}

In a flat universe, the number of galaxies within a solid angle $\Omega$ with 
magnitude in the range $M_{\min}$--$M_{\max}$ and comoving distance $D_{\rm 
com}$ in the range $D_1$--$D_2$ is calculated by
integral:
\begin{eqnarray}
	N=\Omega \int_{D_{1}}^{D_{2}}  dD_{\rm com}\,D_{\rm com}^2 
		\int_{M_{\min}}^{M_{\max}} dM\,\Phi(M) \;. \label{number}
\end{eqnarray}

The solid angle $\Omega$ is the solid angle covered by a survey, and
$M_{\max}=M_{\max}(m,z)$ is the maximum absolute magnitude arising from
the apparent magnitude limit $m$ of the telescope. The comoving distance to 
a galaxy at redshift $z$ is calculated by
\begin{eqnarray}
	D_{\rm com}=\frac{c}{H_0} \int_0^z \frac{dz}{\sqrt{\Omega_m\,
		(1+z)^3+\Omega_{\Lambda}}} \;,
\end{eqnarray}
where $c$ is the speed of light, $H_0$ is the Hubble constant, $\Omega_m$ is 
the fraction of mass contained in baryonic and dark matter in the Universe,
and $\Omega_\Lambda$ is the fraction of mass contained in the cosmological
constant or dark energy.

Throughout the paper we adopt a cosmology with $H_0 = 70$ km s$^{-1}$ 
Mpc$^{-1}$, $\Omega_m=0.3$, and $\Omega_\Lambda=0.7$.

\subsection{Morphology and Redshift Dependent LF}

The LF is one of the fundamental observational properties of galaxies, and the 
amount of work dedicated by different groups of people to derive an accurate 
LF is substantial. The LF has been measured from many galaxy surveys with 
differing sample selections and redshift coverage, and different outcomes are
compared by de Lapparent et al. (2003). The results in the literature have 
shown that there is no universal galaxy LF. Instead, the galaxy LF evolves 
with redshift and galaxy morphology.

It has been found that, in general, the faint-end LF of early-type galaxies
is steeper than that of late-type galaxies, and the characteristic luminosity
of early-type galaxies is smaller than that of late-type galaxies (Madgwick et
al. 2002; Nakamura et al. 2003).

The LF of local galaxies is now well constrained by two large spectroscopic 
surveys: the Two-Degree Field Redshift Survey (2dFGRS; e.g. Norberg et al. 
2002), and the Sloan Digital Sky Survey (SDSS; e.g. Blanton et al. 2003). The 
Canada-France Redshift Survey (CFRS), which includes galaxies up to $z\sim 1$,
showed that the LF evolves with the cosmic redshift and the
evolution depends on the galaxy populations. 

For example, the CFRS survey
shows unambiguously that the population evolves and that this evolution is 
strongly differential with color and, less strongly, with luminosity (Lilly 
et al. 1995). The LF of red galaxies changes little over $0.05<z<1$, while the
LF of blue galaxies shows substantial evolution at redshifts $z > 0.5$.

At higher redshift, the evolution of LF in blue bands over the redshift range
$0.5<z<5.0$, and in red bands over the redshift range $0.5<z<3.5$, has
been derived from the FORS Deep Field survey (Gabasch et al. 2004, 2006). The 
LF measurements for different galaxy types have been derived up to 
$z=1.5$ from the VVDS survey (Zucca et al. 2006).

In this paper, we consider the LF for each type of galaxies separately.

\section{The Radius-Luminosity Relation} \label {radius}

To calculate the probability for a GRB to be coincident with a galaxy
on the sky, we need to measure the projected radius of the galaxy on the
sky. To do so, we associate with each galaxy a physical projected area, as 
a function of the redshift and of the luminosity of galaxies.
 
For an elliptical galaxy, we assume that the area covered by the galaxy on 
the sky is $S=\pi R^2$, where $R$ is an averaged radius. For a spiral or an
irregular galaxy, which is not spherical, we assume that the galaxy has a
random distribution in orientation. For a spiral or an irregular galaxy with 
an inclination angle $\theta$, the area on the sky is $S'=\pi R^2 \cos 
\theta$ ($0<\theta<\pi/2$).

Generally, the size of a galaxy is correlated with its luminosity. Hence,
The value of $R$ for a galaxy with a given luminosity can be derived from a 
statistical relation between the observed radius and luminosity, at a given
redshift. The relation can be fitted by a power law
\begin{eqnarray}
	R=\left(\frac{L}{\zeta _i}\right)^{\varphi_i} \;,
	\label{rr} 
\end{eqnarray}
where $R$ is in kpc and $L$ is in erg s$^{-1}$, $\zeta _i$ and $\varphi_i$ 
are constants that depend on the galaxy morphological types and on the 
bandpass of the telescope.

Dahlen et al. (2007) have shown that the galaxy size evolves strongly with 
redshift. In particular they have claimed that there is a similar evolution
in the size-luminosity relation in several wavelengths, over the range $0<z<
6$. The evolution is consistent with the form $R_h \propto (1+z)^\beta$, where 
$\beta \sim -1$ and $R_h$ is the half-light radius of the galaxy. The ratio 
$R_{90}/R_{50}$ (radius containing $90\%$ and $50\%$ of the flux) is 
approximately constant for de Vaucouleurs and exponential 
profile galaxies ($\sim 3.3$ and $\sim 2.3$, respectively), so we 
can assume that $R_{90}\propto (1+z)^\beta$. 

The solid angle occupied by a galaxy is then
\begin{eqnarray}
	\omega _{\rm gal} = \frac{\langle S\rangle}{D_A^2} \;, \label{area}
\end{eqnarray}
where $D_A$ is the angular-diameter distance to the galaxy [related to the 
comoving distance by Etherington's reciprocity law $D_A=D_{\rm com}/
(1+z)$; Etherington 1933], $\langle S\rangle$ is the projected area of the 
galaxy averaged over inclination.

For spherical or elliptical galaxies we have $\langle S\rangle =\pi R_0^2(1+z
)^{2\beta}$, where $R_0$ is the radius of the galaxy. For disk spiral galaxies
and irregular galaxies with a random 
distribution of inclination we have $\langle S\rangle = (1/2) \pi R_0^2
(1+z)^{2\beta}$. 

\section{Computation of the Probability}

The probability for a GRB to be aligned to a galaxy is always small. Hence,
the probability is simply given by the ratio of the solid angle spanned by
galaxies to the total solid angle
\begin{eqnarray}
	P = \frac{\Omega _{\rm gal}}{\Omega} \;, \label{gen}
\end{eqnarray}
where $\Omega$ is the solid angle of space covered by a survey, and 
$\Omega _{\rm gal}$ is the total solid angle occupied by galaxies. 
Using relations (\ref{number}) and (\ref{area}), the total solid angle 
occupied by galaxies is ($\Omega = 4\pi$)
\begin{eqnarray}
	\Omega_{\rm gal} =4\pi \int_{0}^{z_{\max}} dD_{\rm com}\,
		D_{\rm com}^2 \int_{M_{\min}}^{M_{\max}} dM\,\frac{
		\langle S\rangle}{D_A^2} \, \Phi(M) \;, \label{gen2}
\end{eqnarray}
where $z_{\max}$ is the maximum redshift that can be reached by a survey.
Then, by Etherington's reciprocity law, we have
\begin{eqnarray}
	P = \int_{0}^{z_{\max}}dD_{\rm com}\,\int_{M_{\min}}^{M_{\max}} dM\,
		\langle S\rangle (1+z)^2\, \Phi(M) \;,
	\label{gen3}
\end{eqnarray}
we will adopt $\beta=-1$ to compute $\langle S\rangle$.

Since the LF decays exponentially toward the bright end, the exact value
of $M_{\min}$ does not affect the final result. In our numerical calculation
we take $M_{\min} = -30$. For a given luminosity distance $D_{\rm lum}$, the 
value of $M_{\max}$ is related to the magnitude limit of the telescope, $m$,
by
\begin{eqnarray}
	M_{\max} = m -5\log \frac{D_{\rm lum}}{10~{\rm pc}} -K \;,
\end{eqnarray}
where $K$ is the K-correction depending on the filter.

\begin{figure}
{\includegraphics[width=7cm, angle=-90] {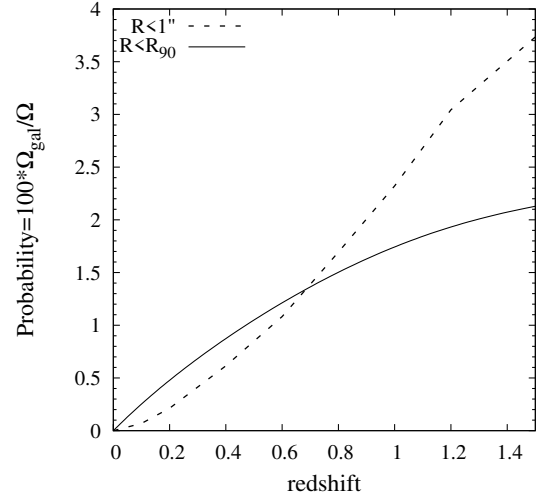}}
\caption{Probability for a GRB to be coincident with a galaxy on 
the sky, with $0<z<1.5$ and the $B$-band magnitude limit $m_b=26.5$. The 
solid curve
is calculated with equation (\ref{gen3}), which assumes that a GRB is
associated with a galaxy if the linear distance from the GRB to the galaxy
center is less than $R_{90}$. The dashed curve is calculated with equation 
(\ref{gen3a}), which assumes that a GRB is associated with a galaxy if the 
angular distance from the GRB to the galaxy center is less than 
$1^{\prime\prime}$.}
\label{probvvds}
\end{figure}

The parameters in the LF are derived from the SDSS and the VVDS catalogs 
(Nakamura et al.2003; Zucca et al. 2006) and the radius-luminosity relation
 from the SDSS catalogs (York et al.2000; Appendixes A \& B). Then,
the probability can be calculated by equation (\ref{gen3}).
The calculated results of $P$ for the parameters in the $B$-band obtained from 
the VVDS 
survey (Zucca et al. 2006) are shown in Fig. \ref{probvvds} (solid line).
In the calculations the K-corrections were provided by E. Zucca (see
also Fukugita, Shimasaku, \& Ichikawa 1995).

For a given galaxy survey, the projected area of resolved galaxies on the
sky can be measured. Then equation (\ref{gen}) can be directly applied to 
calculate the probability for chance coincidence of a GRB with a galaxy on
the sky. As an example,
the fraction of the sky covered by galaxies in the Hubble Deep Fields (HDFs) 
is $\sim 5\%$ if the boundary of a galaxy is defined by twice the isophotal
radius containing $\sim 90\%$ flux (Bernstein, Freedman, \& Madore 2002).

\section{Comparison with observed host galaxies}

\begin{figure}
\centering
{\includegraphics[width=7cm, angle=-90] {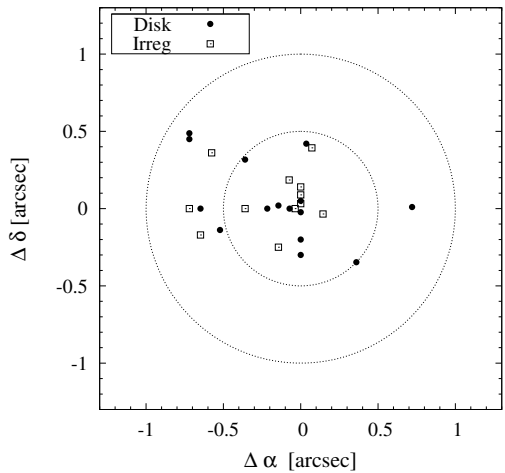}}
\caption{Displacement of GRB positions with respect to the center of
their host galaxies for a sample of 27 long GRBs with well measured and
resolved hosts (in the redshift range $0.089\le z\le 3.42$).}
\label{ra_dec}
\end{figure}

For those GRBs with known hosts, we find out the distribution of the 
distance, projected on the sky, between the GRB's position and the
center of the associated host galaxy.
 From this distribution, we can check 
if the observed distance is within the galaxy radius defined in the equation 
(\ref{rr}). From $\sim 50$ GRBs with both redshift and host associate,
\footnote{http://www.grbhosts.org/} we select 
$27$ long GRBs , including only GRBs with sure host galaxy 
types and R-band magnitudes.
 Figure~\ref{ra_dec} shows the displacement 
between RA/DEC of the GRB's position and the host galaxy counterpart in the 
sample (see table C1 of the Appendix).
 Most GRBs with reliable host measurements have a separation 
smaller than $1^{\prime\prime}$ from the center of their counterpart.
Our comparison is  in agreement with previous works \citep{blo02,fru06}.

\begin{figure}
\centering
{\includegraphics[width=7cm, angle=-90] {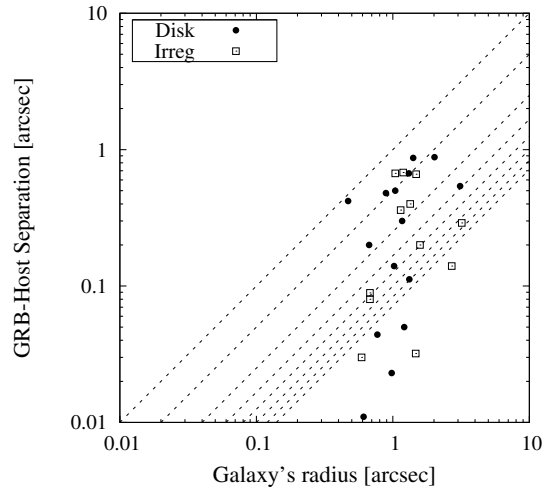}}
\caption{Radius of the host galaxies, computed with equation (\ref{rr}), 
versus the observed distance between GRBs for the same sample and their host 
galaxies. Dashed lines labels the relations of separation/galaxy radius $= 1/14,
1/12$, $1/10$, $1/8$, $1/6$, $1/4$, $1/2$, $1$ (blue-dashed line).}
\label{compa}
\end{figure}

In Fig.~\ref{compa}, we compare the galaxy radius defined by equation 
(\ref{rr}) and the observed distance between GRBs from the center of their
host galaxies.
Since all of the observed GRBs fall in the defined galaxy 
radius, equation (\ref{rr}) provides a reasonable estimate for galaxy radii
and a scale measuring the association of GRBs and their hosts. The probability
calculated with the galaxy radius that we have defined would lead to a 
reasonable estimate on the probability for a GRB to be coincident with a 
galaxy on the sky.

In practice people often identify association of a GRB with a galaxy by
requiring that the projected distance from a GRB to the center of a galaxy
on the sky is smaller than a critical angular distance, say 
$<1^{\prime\prime}$. Then the probability for chance coincidence of a
GRB and a galaxy on the sky is calculated by
\begin{eqnarray}
	P = \omega_{\rm c}\int_{0}^{z_{\max}}D_{\rm com}^2dD_{\rm com}\,
               \int_{M_{\min}}^{M_{\max}} dM\, \Phi(M) \;,
	\label{gen3a}
\end{eqnarray}
where $\omega_{\rm c} = \pi (1^{\prime\prime})^2 = 7.384\times 10^{-11}$
is the solid angle corresponding to a circle of radius $1^{\prime\prime}$.
The probability calculated with this formula is shown in Fig.~\ref{probvvds}
with a dashed curve.

It appears that the probability calculated with an angular radius $1^{\prime
\prime}$ is higher than that calculated with a linear radius $R_{90}$. This
is caused by the fact that for a fixed angular radius the corresponding
linear radius increases with distance while the luminous $R_{90}$ decreases
with redshift. Our results indicate that identifying the GRB-galaxy 
association with a linear distance scale is more reliable than 
with an angular scale from the galaxy center.

\section{Conclusions}

We have calculated the probability for a GRB to be coincident with a galaxy
on the sky, using the luminosity function and the radius-luminosity relation
derived from the SDSS and VVDS surveys. 

Since there is not a reliable 
luminosity function available to higher redshifts, the probability is
calculated only up to a redshift $z\sim 1.5$ (Fig. \ref{probvvds}). The 
results are in agreement with that of \citet{cob06b} and \citet{cob08} which
were obtained with different approaches. The total probability at $z=1.5$
is a few percent. 

We have also calculated the probability of chance coincidence with a
criterion that a GRB is considered to be associated with a galaxy if
the distance from the GRB to the galaxy center is smaller than $1^{\prime
\prime}$ (Fig.~\ref{probvvds}, dashed line). This probability is larger
than that calculated with $R_{90}$ for $z>0.7$ (Fig.~\ref{probvvds}, 
solid line), caused by the fact that
for a fixed angular separation the corresponding linear separation
increases with $z$ and $R_{90}$ decreases with $z$.

Although the chance probability is small, it warns us that identifying 
a GRB host based only on the superposition of a GRB with a galaxy on the
sky is dangerous. So far about 350 GRBs have been detected by {\em Swift},
our results imply that several chance coincidence of a GRB with a galaxy
might have already happened. As a result, some GRB hosts that have been
found might be superficial. However, for the case of GRB 060614, calculation
of the chance superposition of it and a $z<0.125$ galaxy with separation 
$<0.5^{\prime\prime}$ leads to a probability $P = 0.02\%$, consistent
with the result of \citet{gal06}. This small probability indicates that
the association of GRB 060614 and its host is secure.

Obviously, a secure identification of a GRB's host would be obtained by (1) 
the superposition of the GRB with a galaxy; and (2) the afterglow of the GRB
and the host candidate give rise to the same measured redshift.

We have also calculated the probability directly from the data of SDSS,
following the approach of Cobb et al. The results are presented in
Appendixes A, which agree with our analytical results.

\section*{Acknowledgments}
We thank R. Narayan and R. Sunyaev for valuable comments and suggestions, 
E. Zucca for providing the K-corrections for the VVDS survey, and an 
anonymous referee for a very enlightening report.

\begin{appendix}

\section{The SDSS Catalog}

\begin{figure*}
\begin{center}
{\includegraphics [width=5cm, angle =-90] {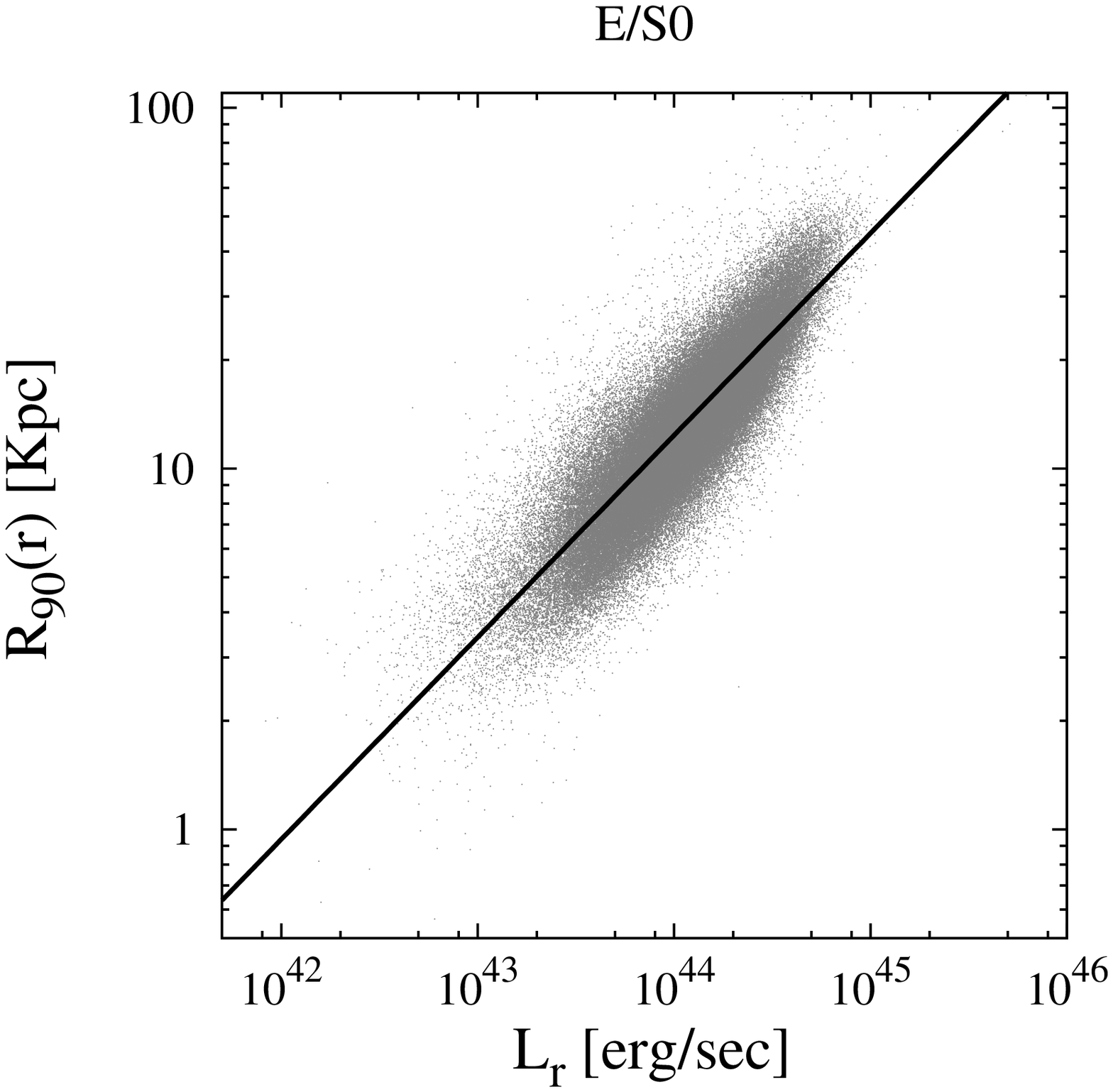}}
{\includegraphics [width=5cm, angle =-90] {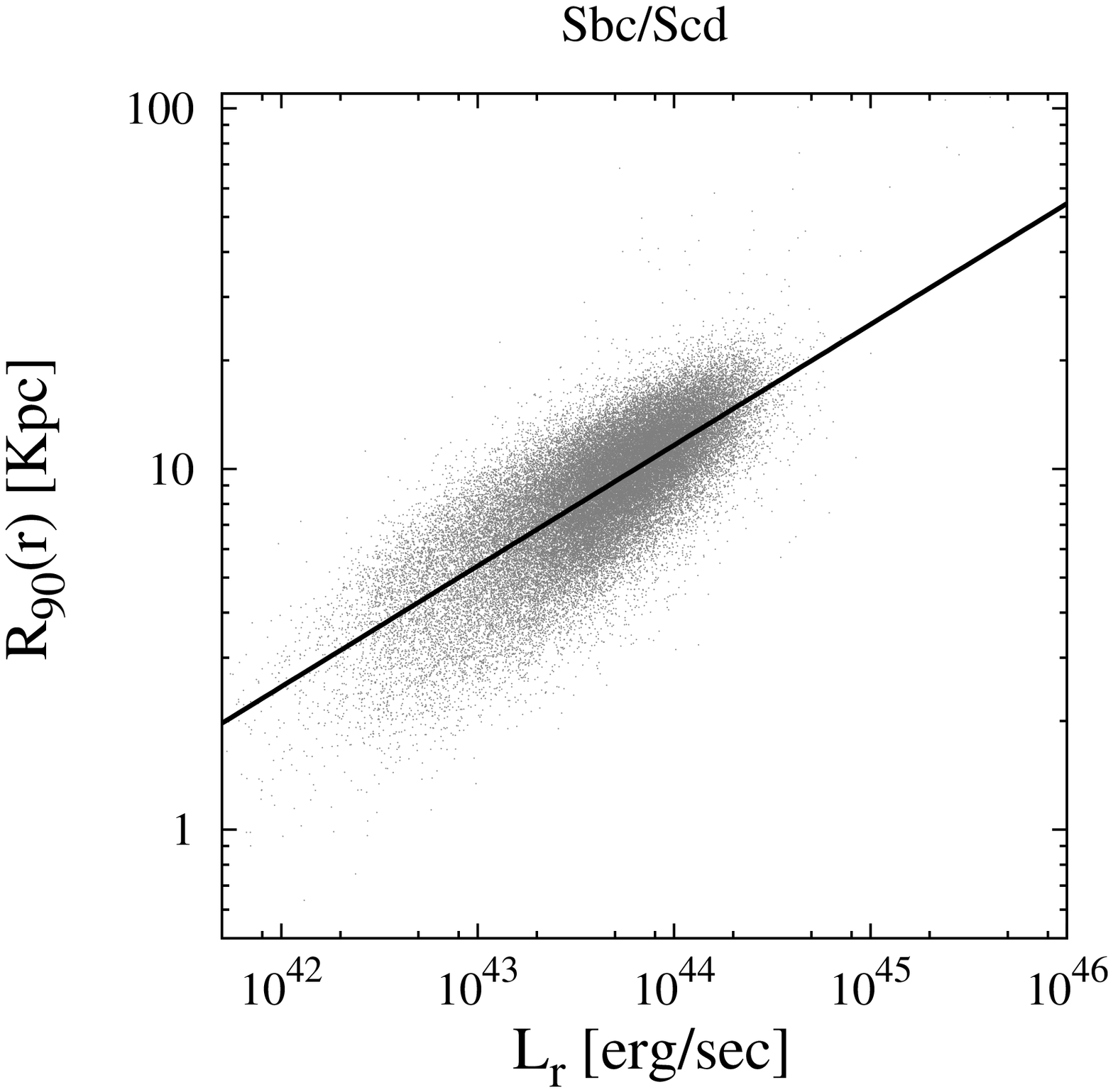}}
{\includegraphics [width=5cm, angle =-90] {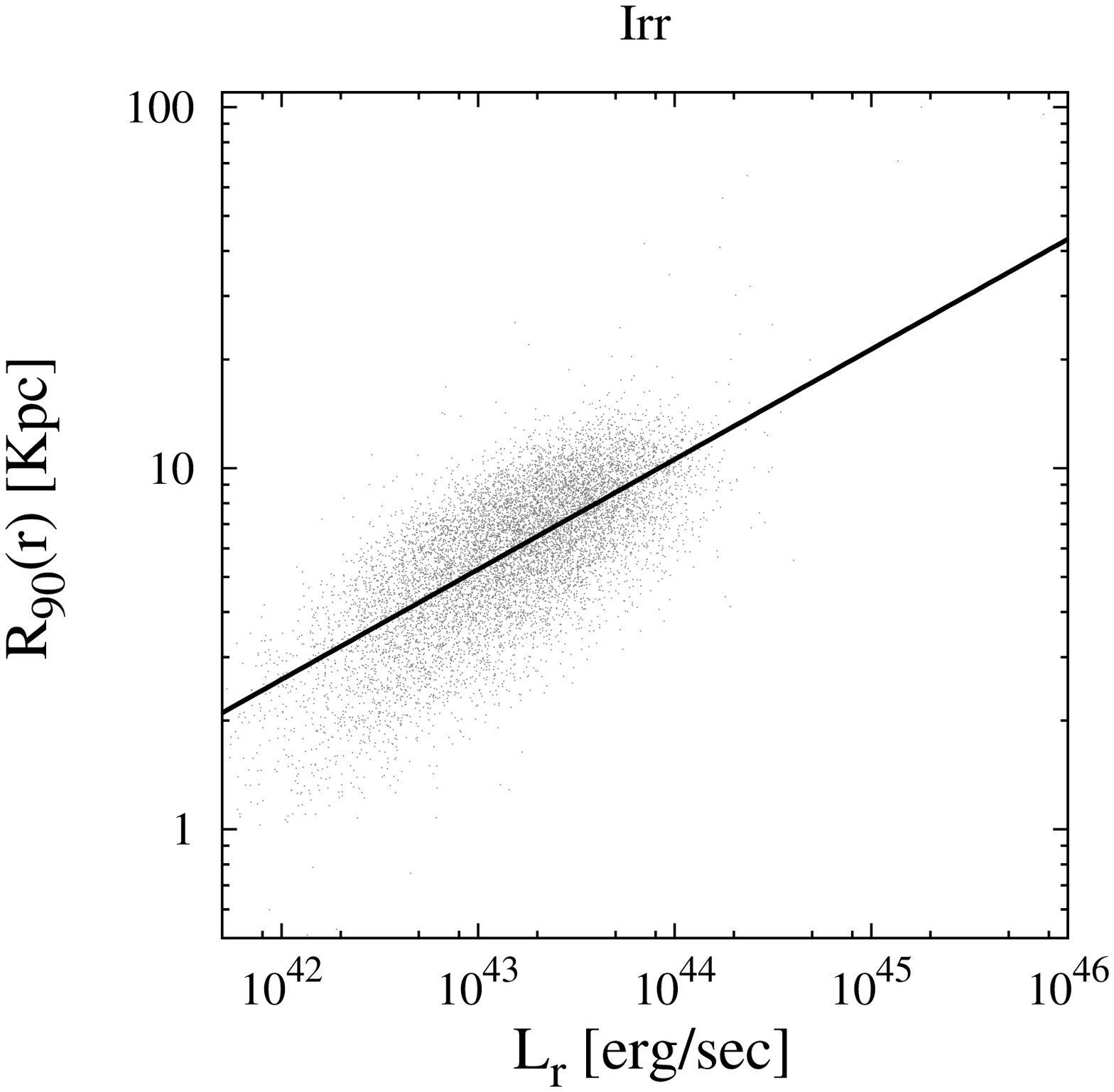}}
\caption{The sky-projected $R_{90}$ versus the luminosity in the $r^*$-band 
for the SDSS galaxies with $15.6<r^*<19.6$ and $0<z<0.3$. Each panel 
corresponds to a galaxy type (elliptical/S0, spiral/Sbc-Scd, and Irregular 
galaxies). The solid line is the best-fit with equation (\ref{rr}) (divided
by $\sqrt{2}$ for spiral and irregular galaxies for the average random 
projection effect). The values of the fitted parameters are listed in 
Table \ref{tablerad}.}
\label{raggi}
\end{center}
\end{figure*}

SDSS catalog (York et al. 2000)\footnote{http://www.sdss.org/} contains the 
largest redshift sample of galaxies with both photometric 
and spectroscopic observations. It is a homogeneous data set that is suitable 
for statistical studies of galaxies. The SDSS sample for LF measurements 
(Nakamura et al. 2003) contains $\sim 1500$ bright galaxies, in the redshift 
range 
$0.01<z<0.12$ with $13.2\leq r^*\leq15.9$. The galaxies are classified into 
four groups by the $g-r$ color: $0\leq T \leq1.0$ (corresponding to Hubble 
type E-S0), $1.5\leq T \leq 3$ (S0/a-Sb), $3.5\leq T \leq5.0$ (Sbc-Sd), 
and $5.5\leq T \leq6$ (Im) (Fukugita et al. 1995). The LF was calculated 
with three methods: ML (maximum likelihood), SWML (stepwise maximum likelihood),
and the $V_{\max}$ method.

To derive the radius-luminosity relation, we take from the SDSS catalog the 
Petrosian radius $R_{90}$ in the $r^*$ band (which contains $90\%$ of the 
Petrosian flux), the value of the apparent brightness $r^*$, and the galaxy
redshift. Note that, SDSS has adopted a modified form of the Petrosian 
system (Petrosian 1976) to define the radius of a source (see Blanton et 
al. 2001 for details). For the exponential profile, $R_{90}$ corresponds to 
the true 90$\%$ light radius, while for the de Vaucouleurs profile $R_{90}=
0.43 R_{90 {\rm true}}$.

Then we convert the apparent magnitude $r^*$ to an absolute magnitude $M$ and 
luminosity $L$, adopting the K-correction supplied by Fukugita et al. (1995) 
for each galaxy type.

The derived sky-projected radius and luminosity in the $r^*$-band for about 
240000 galaxies with $r^*<19.6$ in a sky region of 2500 $\rm deg^2$ are shown
in Fig. \ref{raggi} for different galaxy types (elliptical, spiral, and 
irregular). Power-law fits to the $R$-$L$ relation (eq. \ref{rr}; divided
by $\sqrt{2}$ for spiral and irregular galaxies for the average random 
projection effect) are summarized in
Table \ref{tablerad}.

\begin{table}
\begin{center}
\begin{tabular}{ccc}
\hline
\hline
Galaxy type & ${\zeta _i}$ & $\varphi_i$\\
\hline
E/S0 & 1.12e+42&0.561\\
Sbc/Scd & 2.33e+40&0.335\\
Irr & 1.40e+40&0.305\\
\hline
\hline
\end{tabular}
\caption{Best-fit parameters in the radius-luminosity relation (\ref{rr}) in 
the $r$-band for SDSS galaxies, with $R$ in kpc and $L$ in erg s$^{-1}$.}
\label{tablerad}
\end{center}
\end{table}

\begin{figure}
{\includegraphics [width=7cm, angle =-90] {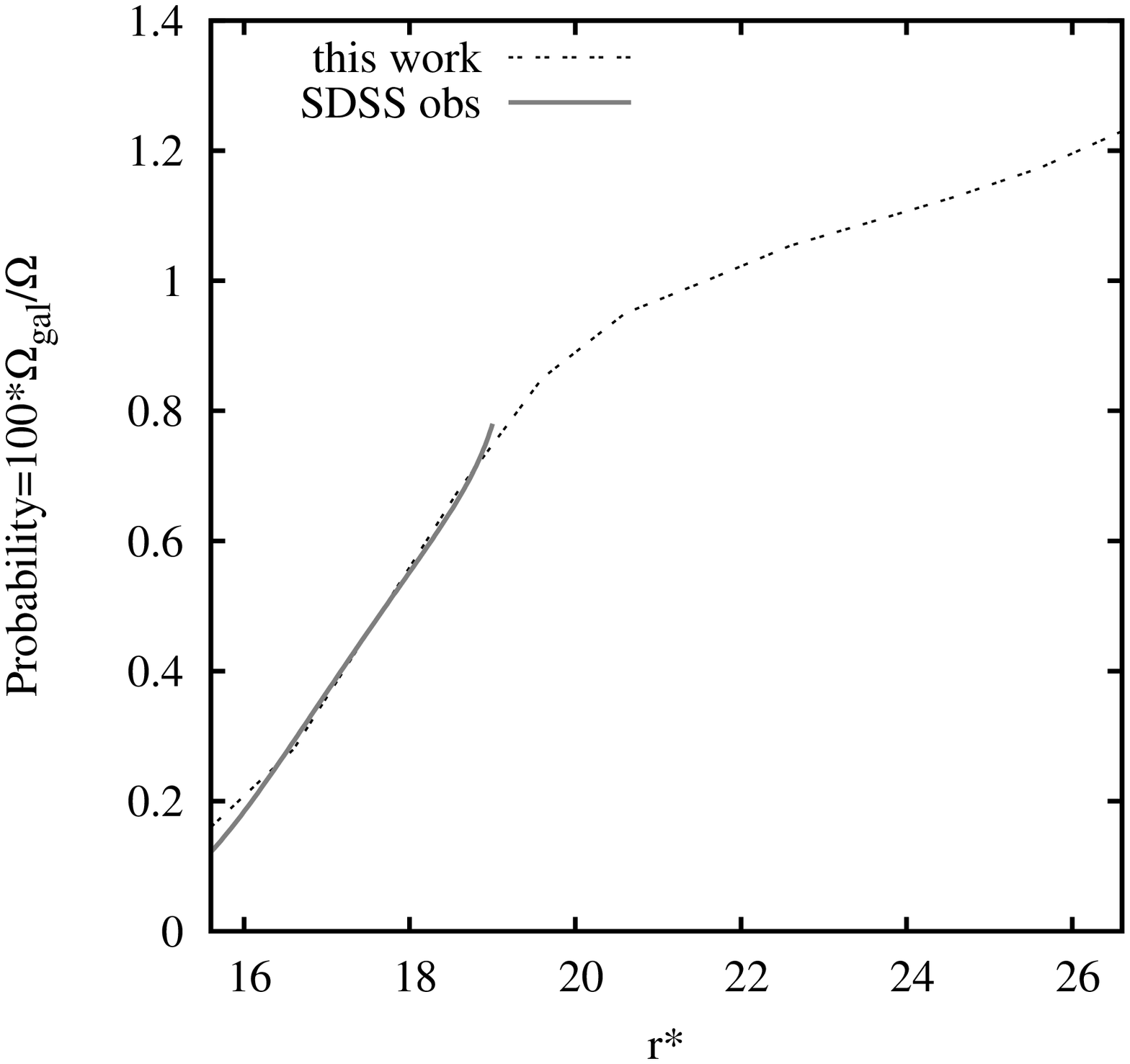}}
\caption{Probability for a GRB to be coincident with a galaxy on 
the sky, obtained from the SDSS galaxies with $0<z<0.3$. The solid 
line is the ratio of the total solid angle occupied  by the SDSS galaxies
to the total solid angle covered by the survey. The dashed line is 
calculated with equation(\ref{gen3}).}
\label{prob}
\end{figure}

We selected a subsample of galaxies with $15.6<r^*<19.6$ (to be consistent
with the magnitude limit adopted in eq. \ref{gen3}), and computed the 
ratio of the sky area covered by them to the total area of the sky. The
results are shown in Fig. \ref{prob}. They agree well with that calculated with
equation (\ref{gen3}) with $z<0.3$ (the redshift covered by the SDSS 
galaxies).

\section{The VVDS LF}\label{vvds}

The VIMOS VLT Deep Survey (VVDS) is a deep spectroscopic survey, containing
galaxies up to redshift $z\sim 1.5$. The first epoch VVDS deep sample covers 
a sky area of about 2200 arcmin$^2$, containing about 7700 galaxies with $17.5
\le I_{AB}\le 24$. Using this sample, Zucca et al. (2006) derived an evolving
and morphology-dependent LF. The galaxies were divided into four types: E/S0,
early spiral, late spiral, and irregular. The parameters in the derived
luminosity functions are listed in table 3 of \citet{zuc06}. The best-fit 
parameters for the radius-luminosity relation are listed in Table 
\ref{radius22}. 

\begin{table}
\begin{center}
\begin{tabular}{ccc}
\hline
\hline
Galaxy type&${\zeta _i}$& $\varphi_i$\\
\hline
E/S0&3.32e+41&0.413\\
Sbc/Scd&3.36e+40&0.322\\
Irr&8.46e+40&0.361\\
\hline
\hline
\end{tabular}
\caption{Best-fit parameters in the radius-luminosity relation (\ref{rr}) in 
the $B$-band  for VVDS galaxies, with $R$ in kpc and $L$ in erg s$^{-1}$.}
\label{radius22}
\end{center}
\end{table}

\section{The Long GRB sample}\label{grb}
In Table C1 we provide details on the long GRBs used in the 
Figs. \ref{ra_dec} and \ref{compa}. The sample contains 27 GRB hosts from 
the {\it{GHostS}} archive (http://www.grbhosts.org). 

\begin{table*}
\begin{minipage}{155mm}
\caption{Sample of 27 long GRBs with known redshift and host.}
\begin{tabular}{ccccccccccc}
\hline
\hline
 GRB&  $z$    &$d_L\,$  &  $M_R$&$R_{90}\,$&$D\,$&$\alpha_H$&$\delta_H\,$&$\alpha_{GRB}\,$&$\delta_{GRB}\,$&Ref.\\
 name	&    &[Mpc]	  &	   &[arcsec]&[arcsec]&(J2000.0)&(J2000.0)&(J2000.0)&(J2000.0)&\\
 (1) &(2)	  &(3)     &(4)      &(5)	  &(6)       &(7)	      &(8)	       &(9)		    &(10) &(11)\\
 \hline
 \multicolumn{11}{l}{Disk galaxies}\\
 \hline
 030323& 3.372 &  29251 &  27.28  &  1.02    & 0.14     &166.5391   & -21.7704      &  166.5391 &  -21.77033      &   [1]\\  
 020819& 0.41  & 2245   & 19.48   &  3.11    & 0.54     &351.8310   &    +6.2655    &     351.831145&   +6.26554  &   [2]  \\ 
 011211& 2.141 & 16888  & 25.97   &  1.04    & 0.49     & 168.8250  &     -21.9489  &     168.8249  &   -21.94894 &   [3]        \\
 011121& 0.362 & 1939   & 23.23   &  2.02    & 0.88     & 173.6240  &     -76.0282  &     173.6235  &   -76.0282  &   [4]        \\
 010921& 0.451 & 2513   & 22.58   &  1.17    & 0.29     & 344.0000  &     +40.9312  &     344.000   &   +40.93128  &   [5] \\
 010222& 1.48  &  10735 &  25.61  &  0.77    & 0.044    &223.0520   & +43.0184      &  223.05202    &  +43.0184       &   [6]\\
 000418& 1.118 &  7594  &  24.15  &  0.98    & 0.023    &186.3300   & +20.1031      &  186.3300     &  +20.10311     &   [7]\\
 000210& 0.846 & 5386   & 24.22   &  0.89    & 0.48     & 29.8149   &     -40.6592  &     29.815    &   -40.6591  &   [8] \\
 991208& 0.706 &  4315  &  24.6   &  0.67    & 0.19     &248.4730   & +46.4558      &  248.4730     &  +46.455834      &   [9]\\
 990712& 0.434 & 2401   & 22.45   &  1.21    & 0.049    & 337.9710  &     -73.4079  &     337.97096 &   -73.40786  &   [10] \\
 990705& 0.842 & 5355   & 22.78   &  1.41    & 0.87     & 77.47670  &     -72.1317  &     77.4770   &   -72.1315  &   [11] \\  
 990123& 1.6   & 11817  & 24.41   &  1.31    & 0.67     & 231.3760  &     +44.7664  &     231.3762  &   +44.7664  &   [12] \\
 980703& 0.966 &  6341  &  22.9   &  1.32    & 0.11     &359.7780   & +8.58530      &  359.77780    &  +8.585300      &   [13]\\
 970508& 0.835 &  5300  &  25.2   &  0.61    & 0.011    &103.4560   & +79.2721      &  103.45604    &  +79.27208      &   [14]\\
 970228& 0.695 & 4233   & 25.88   &  0.47    & 0.43     &  75.4444  &     +11.7816  &     75.4442  &   +11.78159  &   [15] \\
 \hline
 \multicolumn{11}{l}{Irregular galaxies}\\
 \hline 		       
 060614& 0.125 &587   &22.52  &   1.14 & 0.36 &320.8839 &   -53.0267&  320.884	  &  -53.0267  &  [16]  \\ 
 060218& 0.0331&148   &20.16  &   3.21 & 0.29 &50.4153  &   +16.8671&  50.41534   &  16.86717  &  [17] \\
 050826& 0.296 &1535  &21.67  &   1.34 & 0.40  &87.7566  &   -2.6433 &  87.75658   &  -2.64327  &  [18]  \\
 041006& 0.712 &4360  &25.15  &   0.59 & 0.03 &13.7093  &   +1.2349 &  13.70931   &  +1.23490  &  [19]  \\
 030528& 0.782 &4890  &22.0   &   1.48 & 0.66 &256.0010 &   -22.6194&  256.0012   &  -22.6194  &  [20]  \\ 
 030429& 2.66  &21986 &26.3   &   1.04 & 0.67 &183.2810 &   -20.9138&  183.28118  &  -20.91381 &  [21]  \\
 030328& 1.52  &11110 &24.06  &   1.19 & 0.68 &182.7015 &   -9.3476 &  182.70166  &  -9.34750  &  [22]  \\
 020405& 0.691 &4204  &21.59. &   1.58 & 0.19  &209.5130 &   -31.3728&  209.5130208&  -31.37275 &  [23]  \\
 000926& 2.036 &15882 &24.18  &   1.47 & 0.032&256.0400 &   +51.7862&  256.0400   &  +51.78611 &  [24]  \\
 000911& 1.06  &7094  &25.27  &   0.68 & 0.079 &34.6432  &   +7.7410 &  34.64316   &  +7.74102  &  [25]  \\ 
 980613& 1.097 &7418  &25.33  &   0.68 & 0.089&154.4910 &   +71.4571&  154.4910   &  +71.457083&  [26]  \\
 971214& 3.42  &29750 &26.35  &   2.72 & 0.14 &179.1100 &   +65.2001&  179.1100   &  +65.20013 &  [27]  \\
\hline	
\end{tabular}
\tablecomments{Col. (1) GRB name. Col. (2-3) Redshift and luminosity distance 
in Mpc. Col. (4) Observed host magnitude in the $R$-band AB system 
(Fruchter et al. 2006, and references therein). Col. (5) Radius in arcsec 
used in this work. Col. (6) Distance in arcsec between GRB and host. Col. (7-8)
Positions of the host. Col.(9-10) Positions of the GRB. Col. (11) References: 
[1] \cite{gra03,vre04};
[2] \cite{jak05};
[3]  \cite{gra01,hol02};
[4] \cite{sub01,blo02};
[5]  \cite{pri02};
[6] \cite{hen01,gal03};
[7] \cite{mir00,blo03};
[8] \cite{gor03};
[9] \cite{jen99,chr04} ;
[10]  \cite{sah00,chr04};
[11]  \cite{pal99,lef02};
[12]   \cite{blo99};
[13]  \cite{tay98, djo98,chr04}; 
[14]  \cite{fra97, blo98,chr04}; 
[15]  \cite{mar97, blo01};
[16]  \cite{par06, gal06,geh06};
[17]  \cite{mir06};
[18]  \cite{mir07};
[19]  \cite{yam04,wai07};
[20] \cite{but04, rau04};
[21] \cite{jak04};
[22]  \cite{pet03,gor05};
[23]  \cite{wai07};
[24]  \cite{pri01, fyn01};
[25] \cite{pri00,mas05};
[26] \cite{hjo02, hal98};
[27] \cite{ode98}.
}
\label{GRBstab}
\end{minipage}
\end{table*}

\end{appendix}

\label{lastpage}


\begin{thebibliography}{99}

\bibitem[\protect\citeauthoryear{Berger}{2006}]{ber06}
	Berger E., 2006, in Holt S., Gehrels N., Nousek J., eds, Gamma-Ray 
        Bursts in the Swift Era. Am. Inst. Phys., New York, p. 33

\bibitem[\protect\citeauthoryear{Bernstein et al.}{2002}]{bern02}
	Bernstein R. A., Freedman W. L., Madore B. F., 2002, ApJ, 571, 56

\bibitem[\protect\citeauthoryear{Blanton et al.}{2003}]{bla03}
	Blanton M.~R., et al., 2003, \apj, 592, 819 

\bibitem[\protect\citeauthoryear{Blanton et al.}{2006}]{bla06}
	Blanton M.~R., et al., 2001, \aj, 121, 2358 

\bibitem[\protect\citeauthoryear{Bloom et al.}{2003}]{blo03}	
	Bloom J.~S., Berger E., Kulkarni S.~R., Djorgovski S.~G., Frail 
        D.~A., 2003, \aj, 125, 999 

\bibitem[\protect\citeauthoryear{Bloom et al.}{2002}]{blo02}	
	Bloom J.~S., Kulkarni S.~R., Djorgovski S.~G., 2002, \aj, 123, 1111 

\bibitem[\protect\citeauthoryear{Bloom et al.}{2002b}]{blo02b}	
	Bloom J.~S. et al., 2002, \apjl, 572, L45 

\bibitem[\protect\citeauthoryear{Bloom et al.}{2001}]{blo01}
	Bloom J.~S., Djorgovski S.~G., Kulkarni S.~R., 2001, \apj, 554, 678 

\bibitem[\protect\citeauthoryear{Bloom et al.}{1999}]{blo99}
	Bloom J.~S. et al., 1999, \apjl, 518, L1 

\bibitem[\protect\citeauthoryear{Bloom et al.}{1999b}]{blo99b}
	Bloom J.~S. et al., 1999, GCN 206

\bibitem[\protect\citeauthoryear{Bloom et al.}{1998}]{blo98}
	Bloom J.~S., Djorgovski S.~G., Kulkarni S.~R., Frail D.~A., 1998, 
        \apjl, 507, L25 

\bibitem[\protect\citeauthoryear{Butler et al.}{2004}]{but04}
	Butler N. et al., 2004, in Fenimore E. E., Galassi M., eds, 
        Gamma-Ray Bursts: 30 Years of Discovery. Am. Inst. Phys., New York,
        p. 111 

\bibitem[\protect\citeauthoryear{Cobb \& Bailyn}{2007}]{cob08}
	Cobb B.~E., Bailyn C.~D., 2008, \apj, 677, 1157 

\bibitem[\protect\citeauthoryear{Cobb et al.}{2006b}]{cob06b}
	Cobb B.~E., Bailyn C.~D., van Dokkum P.~G., Natarajan P., 2006, 
        \apjl, 651, L85 

\bibitem[\protect\citeauthoryear{Conselice et al.}{2005}]{con05}
	Conselice C.~J. et al., 2005, \apj, 633, 29 

\bibitem[\protect\citeauthoryear{Christensen et al.}{2004}]{chr04}
	Christensen L., Hjorth J., Gorosabel J., 2004, \aap, 425, 913 

\bibitem[\protect\citeauthoryear{Dahlen et al.}{2007}]{dah07}
	 Dahlen T., Mobasher B., Dickinson M., Ferguson H.~C., Giavalisco M., 
         Kretchmer C., Ravindranath S., 2007, \apj, 654, 172 

\bibitem[\protect\citeauthoryear{de Lapparent et al.}{2003}]{deL03}
	de Lapparent V., Galaz G., Bardelli S., Arnouts S., 2003, \aap, 
        404, 831 

\bibitem[\protect\citeauthoryear{Della Valle et al.}{2006}]{del06}
	Della Valle M. et al., 2006, \nat, 444, 1050 

\bibitem[\protect\citeauthoryear{Djorgovski et al.}{1998}]{djo98}
	Djorgovski S.~G., Kulkarni S.~R., Bloom J.~S., Goodrich R., 
        Frail D.~A., Piro L., Palazzi E., 1998, \apjl, 508, L17 

\bibitem[\protect\citeauthoryear{Etherington}{1933}]{Eth33}
	Etherington I.~M.~H., 1933, Phil. Mag., 15, 761 

\bibitem[\protect\citeauthoryear{Frail et al.}{1997}]{fra97}
	Frail D.~A., Kulkarni S.~R., Nicastro L., Feroci M., Taylor G.~B., 
        1997, \nat, 389, 261 

\bibitem[\protect\citeauthoryear{Fruchter et al.}{2006}]{fru06}
	Fruchter A.~S. et al., 2006, \nat, 441, 463 

\bibitem[\protect\citeauthoryear{Fukugita et al.}{1995}]{fuk95}
	Fukugita M., Shimasaku K., Ichikawa T., 1995, \pasp, 107, 945 

\bibitem[\protect\citeauthoryear{Fynbo et al.}{2006}]{fyn06}
	Fynbo J.~P.~U. et al., 2006, \nat, 444, 1047 

\bibitem[\protect\citeauthoryear{Fynbo et al.}{2001}]{fyn01}
	Fynbo J.~U. et al., 2001, \aap, 373, 796 

\bibitem[\protect\citeauthoryear{Gabasch et al.}{2006}]{gab06}
	Gabasch A. et al., 2006, \aap, 448, 101 

\bibitem[\protect\citeauthoryear{Gabasch et al.}{2004}]{gab04}
	Gabasch A. et al., 2004, \aap, 421, 41 

\bibitem[\protect\citeauthoryear{Galama et al.}{2003}]{gal03}
	Galama T.~J. et al., 2003, \apj, 587, 135 

\bibitem[\protect\citeauthoryear{Galama et al.}{1998}]{gal98}
	Galama T.~J. et al., 1998, \nat, 395, 670 

\bibitem[\protect\citeauthoryear{Gal-Yam et al.}{2006}]{gal06}
	Gal-Yam A. et al., 2006, \nat, 444, 1053 

\bibitem[\protect\citeauthoryear{Gehrels et al.}{2006}]{geh06}
	Gehrels N. et al., 2006, \nat, 444, 1044 

\bibitem[\protect\citeauthoryear{Graziani et al.}{2003}]{gra03}
	Graziani C. et al., 2003, GCN 1956 

\bibitem[\protect\citeauthoryear{Grav et al.}{2001}]{gra01}
	Grav T. et al., 2001, GCN 1191

\bibitem[\protect\citeauthoryear{Gorosabel et al.}{2005}]{gor05}
	Gorosabel J., Jel{\'{\i}}nek M., de Ugarte Postigo A., Guziy S., 
        Castro-Tirado A.~J., 2005, Il Nuovo Cimento C, 28, 677 

\bibitem[\protect\citeauthoryear{Gorosabel et al.}{2003}]{gor03}
	Gorosabel J. et al., 2003, \aap, 400, 127 

\bibitem[\protect\citeauthoryear{Halpern \& Fesen}{1998}]{hal98}
	Halpern J.~P., Fesen R., 1998, GCN 134

\bibitem[\protect\citeauthoryear{Henden et al.}{2001}]{hen01}
	Henden A., Vrba F., 2001, GCN 967

\bibitem[\protect\citeauthoryear{Hjorth}{2002}]{hjo02}
	Hjorth J., et al., 2002, \apj, 576, 113 

\bibitem[\protect\citeauthoryear{Holland et al.}{2002}]{hol02}
	Holland S.~T. et al., 2002, \aj, 124, 639 

\bibitem[\protect\citeauthoryear{Jensen}{1999}]{jen99}
	Jensen B.~L., Hjorth J., Pedersen H., Kristen H.~E., Tomassi L., 
        Pian E., Hurley K., 1999, GCN 454 

\bibitem[\protect\citeauthoryear{Jakobsson}{2005}]{jak05}
	Jakobsson P. et al., 2005, \apj, 629, 45 

\bibitem[\protect\citeauthoryear{Jakobsson}{2004}]{jak04}
	Jakobsson P. et al., 2004, \aap, 427, 785 

\bibitem[\protect\citeauthoryear{Jakobsson}{2003}]{jak03}
	Jakobsson P. et al., 2003, \aap, 408, 941 

\bibitem[\protect\citeauthoryear{King et al.}{2007}]{kin07}	
	King L., Corless V., 2007, \mnras, 374, L37 

\bibitem[\protect\citeauthoryear{Levan et al.}{2007}]{leva07}
	Levan A. J. et al. 2007, \mnras, 378, 1143

\bibitem[\protect\citeauthoryear{Levesque \& Kewley}{2007}]{lev07}
	Levesque E.~M., Kewley L.~J., 2007, \apjl, 667, L121 

\bibitem[\protect\citeauthoryear{Le Floc'h et al.}{2002}]{lef02}
	Le Floc'h E. et al., 2002, \apjl, 581, L81 

\bibitem[\protect\citeauthoryear{Li}{2006}]{li06}
	Li L.-X., 2006, \mnras, 372, 1357 

\bibitem[\protect\citeauthoryear{Li \& Paczy\'nski}{1998}]{li98}
	Li L.-X., Paczy{\'n}ski B., 1998, \apjl, 507, L59 

\bibitem[\protect\citeauthoryear{Lilly et al.}{1995}]{lil95}
	Lilly S.~J., Tresse L., Hammer F., Crampton D., Le Fevre O., 
        1995, \apj, 455, 108 

\bibitem[\protect\citeauthoryear{Loeb}{2002}]{loe02}
	Loeb A., 2002, in Gilfanov M., Sunyaev R., Churazov E., eds,
	Lighthouses of the Universe: The Most Luminous Celestial Objects 
	and Their Use for Cosmology. Springer-Verlag, Berlin, p. 137

\bibitem[\protect\citeauthoryear{MacFadyen et al.}{2001}]{mac01}
	MacFadyen A.~I., Woosley S.~E., Heger A., 2001, \apj, 550, 410 

\bibitem[\protect\citeauthoryear{MacFadyen \& Woosley}{1999}]{mac99}
	MacFadyen A.~I., Woosley S.~E., 1999, \apj, 524, 262 

\bibitem[\protect\citeauthoryear{Madgwick et al.}{2002}]{mad02}
	Madgwick D.~S. et al., 2002, \mnras, 333, 133 

\bibitem[\protect\citeauthoryear{Mangano et al.}{2007}]{man07}
	 Mangano V. et al., 2007, \aap, 470, 105 

\bibitem[\protect\citeauthoryear{Margon et al.}{1997}]{mar97}
	Margon B., Deutsch E.~W., Lamb D.~Q., Castander F.~J., Metzger M., 
        1997, IAUC 6618

\bibitem[\protect\citeauthoryear{Masetti et al.}{2005}]{mas05}
	Masetti N. et al., 2005, \aap, 438, 841 

\bibitem[\protect\citeauthoryear{McBreen et al.}{2008}]{mcb08}
	McBreen S. et al., 2008, \apjl, 677, L85 

\bibitem[\protect\citeauthoryear{Mirabal et al.}{2007}]{mir07}
	Mirabal N., Halpern J.~P., O'Brien P.~T., 2007, \apjl, 661, L127 

\bibitem[\protect\citeauthoryear{Mirabal et al.}{2006}]{mir06}
	Mirabal N., Halpern J.~P., An D., Thorstensen J.~R., Terndrup D.~M., 
        2006, \apjl, 643, L99 

\bibitem[\protect\citeauthoryear{Mirabal et al.}{2000}]{mir00}
	Mirabal N., Halpern J.~P., Wagner R.~M., 2000, GCN 650 

\bibitem[\protect\citeauthoryear{Nakamura et al.}{2003}]{nak03}
	Nakamura O., Fukugita M., Yasuda N., Loveday J., Brinkmann J., 
        Schneider D.~P., Shimasaku K., SubbaRao M., 2003, \aj, 125, 1682 

\bibitem[\protect\citeauthoryear{Norberg et al.}{2002}]{nor02}
	Norberg P. et al., 2002, \mnras, 336, 907 

\bibitem[\protect\citeauthoryear{Odewahn et al.}{1998}]{ode98}
	Odewahn S.~C. et al., 1998, \apjl, 509, L5 

\bibitem[\protect\citeauthoryear{Ofek et al.}{2006}]{ofe06}
	Ofek E.~O. et al., 2007, \apj, 662, 1129 

\bibitem[\protect\citeauthoryear{O'Shaughnessy et al.}{2007}]{osh07}
	O'Shaughnessy R., Kalogera V., Belczynski K., 2007, \apj, 667, 1048 

\bibitem[\protect\citeauthoryear{Paczy\'nski}{1998}]{pac98}
	Paczynski B., 1998, \apjl, 494, L45 

\bibitem[\protect\citeauthoryear{Palazzi et al.}{1999}]{pal99}
	Palazzi E., Masetti N., Pian E., 1999, GCN 382

\bibitem[\protect\citeauthoryear{Parsons et al.}{2006}]{par06}
	Parsons A.~M. et al., 2006, GN 5252 

\bibitem[\protect\citeauthoryear{Peterson \& Price }{2003}]{pet03}
	Peterson B.~A., Price P.~A., 2003, GCN 1974 

\bibitem[\protect\citeauthoryear{Petrosian et al.}{1976}]{pet76}
	Petrosian V. 1976, \apjl, 209, L1 

\bibitem[\protect\citeauthoryear{Price et al.}{2002}]{pri02}
	Price P.~A. et al., 2002, \apjl, 571, L121 

\bibitem[\protect\citeauthoryear{Price et al.}{2001}]{pri01}
	Price P.~A. et al., 2001, \apjl, 549, L7 

\bibitem[\protect\citeauthoryear{Price}{2000}]{pri00}
	Price P.~A., 2000, GCN 799

\bibitem[\protect\citeauthoryear{Rau et al.}{2004}]{rau04}
	Rau A. et al., 2004, \aap, 427, 815 

\bibitem[\protect\citeauthoryear{Sahu et al.}{2000}]{sah00}
	Sahu K.~C. et al., 2000, \apj, 540, 74 

\bibitem[\protect\citeauthoryear{Schechter}{1976}]{sch76}
	Schechter P., 1976, \apj, 203, 297 

\bibitem[\protect\citeauthoryear{Subrahmanyan et al.}{2001}]{sub01}
	Subrahmanyan R., Kulkarni S.~R., Berger E., Frail D.~A., 2001, 
        GCN 1156

\bibitem[\protect\citeauthoryear{Tanvir \& Levan}{2007}]{tan07}
	Tanvir N. R., Levan A. J., 2007, to appear in the proceedings of 
	``070228: The next decade of GRB afterglows'', Amsterdam, March 
	2007 (arXiv:0709.0861)

\bibitem[\protect\citeauthoryear{Taylor et al.}{1998}]{tay98}
	Taylor G.~B., Frail D.~A., Beasley A.~J., Kulkarni S.~R., 1998, 
        GCN 152 

\bibitem[\protect\citeauthoryear{Th\"one et al.}{2008}]{tho08}
 	Th{\"o}ne C.~C. et al., 2008, \apj, 676, 1151 

\bibitem[\protect\citeauthoryear{Vreeswijk et al.}{2004}]{vre04}
	Vreeswijk P.~M. et al., 2004, \aap, 419, 927 

\bibitem[\protect\citeauthoryear{Wainwrigh et al.}{2007}]{wai07}
	Wainwright C., Berger E., Penprase B.~E., 2007, \apj, 657, 367 

\bibitem[\protect\citeauthoryear{Watson et al.}{2007}]{wat07}
	Watson D., Fynbo J.~P.~U., Th{\"o}ne C.~C., Sollerman J., 2007, 
        Phil. Trans. R. Soc. A, 365, 1269 

\bibitem[\protect\citeauthoryear{Woosley \& Bloom}{2006}]{woo06a}
	Woosley S.~E., Bloom J.~S., 2006, \araa, 44, 507 

\bibitem[\protect\citeauthoryear{Woosley \& Heger}{2006}]{woo06b}
	Woosley S.~E., Heger A., 2006, \apj, 637, 914 

\bibitem[\protect\citeauthoryear{Yamaoka et al.}{2004}]{yam04}
	Yamaoka H., Ayani K., Itagaki K., 2004, GCN 2781 

\bibitem[\protect\citeauthoryear{York et al.}{2000}]{yor00}
	York D.~G. et al., 2000, \aj, 120, 1579 

\bibitem[\protect\citeauthoryear{Zhang et al.}{2007}]{zha07}
	Zhang B., Zhang B.-B., Liang E.-W., Gehrels N., Burrows D. N., 
	M\'esz\'aros P., 2007, ApJ, 655, L25

\bibitem[\protect\citeauthoryear{Zhang}{2006}]{zha06}
	Zhang B., 2006, \nat, 444, 1010 

\bibitem[\protect\citeauthoryear{Zucca et al.}{2006}]{zuc06}
	Zucca E. et al., 2006, \aap, 455, 879 

\end{thebibliography}
\end{document}